\documentclass[12pt]{article}
\usepackage{amsmath,amsfonts,amssymb}
\usepackage{geometry}
\usepackage{graphicx}
\usepackage{natbib}
\usepackage{hyperref}
\usepackage{appendix}
\usepackage{subfig}
\usepackage{booktabs}
\usepackage{array}
\usepackage{multirow}
\usepackage{longtable}
\usepackage{titlesec}
\usepackage{setspace}
\usepackage{authblk} 
\usepackage{lipsum}
\usepackage{natbib}
\usepackage{pdfpages}

\titleformat{\section}{\normalfont\Large\bfseries}{\thesection.}{0.5em}{}
\titleformat{\subsection}{\normalfont\large\bfseries}{\thesubsection.}{0.5em}{}

\makeatletter
\def\@fnsymbol#1{\ifcase#1\or *\or \dag\or \ddag\or \S\or \P\or \| \or \#\fi}
\makeatother

\title{\textbf{Advances in Approximate Bayesian Inference for Models in Epidemiology}}

\author[a,b]{Xiahui Li\thanks{Corresponding author at: School of Mathematics and Statistics, University of St Andrews, UK. \\ E-mail address: xl94@st-andrews.ac.uk}}
\author[a,b]{Fergus Chadwick}
\author[a,b]{Ben Swallow}

\affil[a]{\small School of Mathematics and Statistics, University of St Andrews, UK}
\affil[b]{\small Centre for Research into Ecological and Environmental Modelling, University of St Andrews, UK}

\date{}

\begin{document}

\maketitle

\newpage
\section*{Highlights}
\begin{itemize}
    \item Complex epidemiological models are notoriously challenging to fit to data, particularly when real-time updates are required.
    \item We review the main four families of approximate Bayesian methods that allow the user to trade-off precision and uncertainty with computational efficiency.
    \item In particular, we provide a synthesis of Approximate Bayesian Computation (ABC), Bayesian Synthetic Likelihood (BSL), Integrated Nested Laplace Approximation (INLA), and Variational Inference (VI), summarising recent developments and areas of active research in these fields.
    \item We provide a decision-making framework to allow non-specialists to choose the most appropriate framework for their modelling problem.
    \item We conclude by identifying two exciting research frontiers that arise from this synthesis: (1) developing hybrid Bayesian inference methods that strategically integrate the strengths of both exact and approximate techniques, to achieve scalable yet theoretically grounded inference; and (2) applying these advances to answer important epidemiological questions, meeting the growing need for accurate and efficient inference during public health crises.
\end{itemize}
\vspace{2em}

\newpage
\begin{abstract}

Bayesian inference methods are useful in infectious diseases modeling due to their capability to propagate uncertainty, manage sparse data, incorporate latent structures, and address high-dimensional parameter spaces. However, parameter inference through assimilation of observational data in these models remains challenging. While asymptotically exact Bayesian methods offer theoretical guarantees for accurate inference, they can be computationally demanding and impractical for real-time outbreak analysis. This review synthesizes recent advances in approximate Bayesian inference methods that aim to balance inferential accuracy with scalability. We focus on four prominent families: Approximate Bayesian Computation, Bayesian Synthetic Likelihood, Integrated Nested Laplace Approximation, and Variational Inference. For each method, we evaluate its relevance to epidemiological applications, emphasizing innovations that improve both computational efficiency and inference accuracy. We also offer practical guidance on method selection across a range of modeling scenarios. Finally, we identify hybrid exact approximate inference as a promising frontier that combines methodological rigor with the scalability needed for the response to outbreaks. This review provides epidemiologists with a conceptual framework to navigate the trade-off between statistical accuracy and computational feasibility in contemporary disease modeling.

\vspace{1em} 

\noindent \textbf{Keywords:} approximate Bayesian inference; Approximate Bayesian Computation; Synthetic Likelihood; INLA; Variational Inference; calibration; compartmental models; epidemiology; infectious disease models
\end{abstract}

\newpage
\section{Introduction}

The spread of infectious diseases poses significant challenges to public health, economic stability, and societal well-being, as demonstrated by outbreaks such as COVID-19 \citep{hossain2020epidemiology,kaye2021economic}. Mechanistic models are critical tools for quantifying transmission dynamics, predicting outbreaks, and evaluating interventions \citep{brauer2008compartmental}. However, inferring parameters for these models remains challenging. High-dimensional parameter spaces, latent variables (e.g., unobserved infections/infection times), and uncertainties in model structure (e.g., mixing patterns, model assumptions) complicate the estimation process \citep{swallow2022challenges}. These challenges are further exacerbated by incomplete or noisy data, such as inconsistent medical records or delayed test results, which are particularly problematic in real-time applications \citep{marion2022modelling, kretzschmar2022challenges}. While exact methods, like MCMC, remain the gold standard, their computational costs are often prohibitive for complex models or emerging outbreaks, where rapid inference is critical for public health decision-making. Such challenges have spurred interest in approximate Bayesian methods that trade asymptotic exactness for scalability.

This review focuses on recent advances in approximate Bayesian inference for epidemiological models, with a focus on innovations that address three key needs: (1) computational efficiency for high-dimensional or latent variable models, (2) robustness to noisy or sparse data, and (3) integration with statistical and machine learning methods to automate traditionally manual tuning steps. Our goal is to provide a comparative and application-oriented review for epidemiological modelers, particularly those interested in using or developing machine learning tools for inference. We focus on four major approximate Bayesian inference approaches that are gaining traction in the infectious disease modeling community: Approximate Bayesian Computation (ABC), which bypasses likelihood evaluation via simulation-based comparisons; Bayesian Synthetic Likelihood (BSL), which uses Gaussian approximations of summary statistics; Integrated Nested Laplace Approximation (INLA), which uses latent Gaussian models for fast deterministic inference; and Variational Inference (VI), which optimizes parametric approximations to the posterior.

The remainder of this paper is organized as follows: Section 2 reviews exact Bayesian inference methods, evaluating their strengths and limitations for infectious disease modeling applications. In Section 3, we provide a comparative review of the four key approximate methods (ABC, BSL, INLA, and VI), examining their theoretical foundations, implementation challenges, and epidemiological use cases while offering practical method selection guidelines. In Section 4 we conclude by identifying open challenges and future research directions at the intersection of machine learning and epidemiological inference.

By linking methodological advances to applied needs, this review aims to equip researchers with the tools to navigate the trade-offs between accuracy, scalability, and robustness in epidemic inference.

\section{Asymptotically Exact Bayesian Inference Methods}

\subsection{Bayesian Inference in Epidemiological Modeling}

Bayesian inference provides a powerful statistical framework for infectious disease modeling, due to their capability to propagate uncertainty, manage sparse data, incorporate latent structures, and address high-dimensional parameter spaces \citep{lopes2010abc, kypraios2017tutorial, grinsztajn2021bayesian}. Following Bayes' rule, the posterior distribution is defined by $$\pi(\theta|y) = \frac{\pi(y|\theta)\pi(\theta)}{\pi(y)} \propto \pi(y|\theta)\pi(\theta),$$ where $\pi(\theta|y)$ represents our prior beliefs about model parameters $\theta$ before seeing any data $y$, $\pi(y|\theta)$ is the likelihood of observed data $y$. The term $\pi(y) = \int \pi(y|\theta)\pi(\theta)d\theta$, known as the marginal likelihood or evidence, acts like a normalizing constant to ensure that the posterior is a valid probability distribution. In practice, $\pi(y)$ is often unavailable in closed form or requires exponential time to calculate. For this reason, it is common to express the posterior distribution up to a proportionality, omitting $\pi(y)$ while preserving the distribution's shape and position. This formulation still allows for direct quantification of uncertainty in parameters that are essential for epidemic prediction and planning interventions.

To apply the Bayesian framework to real world problems, three conditions must be appropriately specified: (1) set up appropriate prior distributions; (2) be able to use the posterior distribution for uncertainty inference; and (3) the likelihood needs to be tractable, that is, it should be possible to express and compute it explicitly. However, in practice, the chosen priors and likelihoods often do not yield a closed-form solution for the posterior. This is especially true when the likelihood is complex, difficult to evaluate directly, or lacks a closed-form expression. In such cases, calculating the likelihood may require integrating over hidden variables or accounting for all possible stochastic realizations of an epidemic, a process that can be computationally intensive and often infeasible. As a result, computing the quantities of interest is a numerical problem and is a challenge in itself.

Bayesian inference methods have advanced significantly in addressing these computational hurdles and can be broadly classified into asymptotically exact approaches and approximate inference methods, depending on how they estimate the posterior distribution \citep{alahmadi2020comparison}. Exact methods, such as Markov chain Monte Carlo (MCMC), directly sample from the posterior distribution using algorithms like Metropolis-Hastings or Hamiltonian Monte Carlo (HMC), offering accurate results when the likelihood is tractable and there is sufficient time to run the algorithm to convergence \citep{neal2012mcmc, brooks2011handbook}.

\subsection{Metropolis Hastings Monte Carlo Method}

A foundational method within the MCMC family is the Metropolis-Hastings (MH) algorithm \citep{metropolis1953equation, hastings1970monte}, which has been widely used for parameter inference in infectious disease modeling. The MH algorithm constructs a Markov chain, a sequence of correlated random samples that explore the parameter space, such that its stationary distribution corresponds to the posterior. To assess convergence, it is common to run multiple chains and compare their behavior, if the chains mix well and converge to similar distributions, it indicates that the algorithm has likely reached the target distribution. Once convergence is achieved, the samples drawn from the chain provide a reliable approximation of the true posterior distribution.

At each step, the algorithm proposes a new set of parameter values, $\theta'$, and accepts the proposal with a probability determined by the ratio of posterior densities:
$$
\alpha(\theta \xrightarrow{} \theta') = min(1, \frac{\pi(y|\theta')\pi(\theta')}{\pi(y|\theta)\pi(\theta)}).
$$
If the proposed values lead to a higher posterior density, they are accepted deterministically. If not, the proposal may still be accepted with a probability proportional to how close it is in posterior value to the current state, allowing the chain to explore the parameter space while still favoring higher-probability regions.

While widely used, the MH algorithm often suffers from slow convergence and poor mixing in real-world epidemiological scenarios due to high parameter correlations, multimodal posterior distributions, and increased dimensionality arising from realistic epidemic dynamics \citep{hoffman2014no, neal1993probabilistic}. Such inefficiencies largely arise from their reliance on random-walk proposals, which limits its ability to efficiently explore the posterior landscape \citep{gelman1995bayesian, betancourt2017conceptual}.

\subsection{Hamiltonian Monte Carlo Method}

Hamiltonian Monte Carlo (HMC) was developed to overcome these inefficiencies by integrating concepts from Hamiltonian dynamics \citep{duane1987hybrid, neal2012mcmc} and differential geometry \citep{betancourt2014optimizing}. 
HMC augments the parameter space by introducing auxiliary momentum variables and uses gradient information from the target (i.e. posterior) distribution to guide the sampling process. This approach significantly reduces random-walk behavior, enabling efficient exploration of high-dimensional or highly correlated posterior distributions \citep{neal2012mcmc}. For a detailed technical description of the HMC method, see Appendix A. 

Although HMC significantly improves sampling efficiency over classical MCMC algorithms \citep{monnahan2017faster}, its performance depends on several tuning parameters, including the step size, the number of leapfrog steps, and the covariance structure of the momentum variables \citep{betancourt2016identifying}. These parameters greatly influence the sampler's performance but can be challenging to optimize manually. To address this, modern probabilistic programming tools (e.g., Stan and PyMC) implement HMC along with adaptive variants like the No-U-Turn Sampler (NUTS), which automate parameter tuning, making Bayesian inference more accessible to epidemiological research \citep{hoffman2014no, carpenter2017stan, abril2023pymc}. Recent epidemiological studies have employed HMC through Stan to efficiently infer parameters in a range of infectious disease models \citep{chatzilena2019contemporary, andrade2020evaluation, grinsztajn2021bayesian}.

Despite these advancements, exact Bayesian methods, including HMC, remain computationally intensive when applied to complex, high-dimensional epidemic scenarios. Moreover, exact methods relied heavily on the availability and tractability of likelihood functions, which can be complex or intractable in certain epidemiological settings. 

\section{Approximate Bayesian Inference Methods}

The practical limitations of asymptotically exact inference methods, such as high computational costs and analytical intractability, have driven increased interest and development in approximate Bayesian inference methods. These methods sacrifice some degree of statistical precision, accepting small biases or wider uncertainty intervals, in exchange for significant gains in computational efficiency and scalability. For example, while exact methods aim to sample from the true posterior distribution, approximate approaches often rely on deterministic approximations or surrogate models to achieve faster results. This trade-off is valuable in epidemiology, where timely inference for large datasets or computationally intensive models (e.g., spatial or individual-based simulations) often prioritized over asymptotic exactness. In this section, we review four prominent families of approximate methods: Approximate Bayesian Computation (ABC), Bayesian Synthetic Likelihood (BSL), Integrated Nested Laplace Approximation (INLA), and Variational Inference (VI).

\subsection{Approximate Bayesian Computation}

Many real-world problems in epidemiology involve models with intractable likelihoods, such as agent-based models, stochastic compartmental models, or partially observed transmission trees. In such cases, likelihood-free approaches like Approximate Bayesian Computation (ABC) offer a practical alternative to inference when exact Bayesian inference methods are infeasible.

\subsubsection{Overview of ABC method}

The origins of ABC date back to \citet{rubin1984bayesianly}, with further formalization by \citet{tavare1997inferring} using an acceptance-rejection framework. Since then, ABC has evolved into a diverse family of methods for complex systems. Its core strengths lie in its generality: It bypasses explicit likelihood calculations by comparing summary statistics from observed and simulated data using a distance metric.

The standard rejection-acceptance ABC algorithm (Figure \ref{fig:abcbsl}) operates as follows:

\begin{enumerate}
    \item Sample parameter $\theta^*$ from the prior distribution $\pi(\theta)$.
    \item Simulate data $y_{sim}$ from the model given parameter $\theta^*$.
    \item Reduce observed $y_{obs}$ and simulated data $y_{sim}$ to a set of chosen summary statistics $s(y_{obs})$ and $s(y_{sim})$.
    \item Accept $\theta^*$ if the distance $d(s(y_{obs}),s(y_{sim}))$ is less than a predefined threshold $\epsilon$; otherwise, reject $\theta^*$.
    \item Repeat the process until the desired number of posterior samples is obtained.
\end{enumerate}

The accepted parameters form an approximation to the posterior distribution:
$$
\pi_{ABC}(\theta|s(y_{obs}))\propto \int K_h(d(s(y_{obs}), s(y_{sim})))\pi(s(y_{sim})|\theta)\pi(\theta)d(s(y_{sim}))
$$
where distance $d(s(y_{obs}),s(y_{sim}))$ measures the discrepancy between observed and simulated summaries, $K_h$ is a kernel function that weights smaller distances $d$ more heavily. The distribution is shaped by the choices the practitioner makes for the summary statistics, distance metric, and tolerance level. If we think back to Bayes' rule, the likelihood function is effectively replaced by the simulation process in ABC.
ABC's flexibility makes it well-suited for inference in models with latent variables, non-linear dynamics, or high-dimensional data \citep{beaumont2002approximate}. However, its performance depends on three tuning parameters: summary statistics, the distance metric, and the tolerance level \citep{sisson2018overview, prangle2018summary}. These components will be discussed in detail in Section 3.1.2 and 3.1.3.

\subsubsection{Methodological Advancements and Application}

Over the years, substantial improvements have been made to the classical rejection-acceptance ABC algorithm, particularly through refinements in its three tuning parameters. \citet{harrison2020automatic} introduced a weighted Euclidean distance approach that aim to select the optimal weight vector to maximize the distance between the prior distribution of parameters and the posterior distributions of ABC. In traditional ABC implementations, it is often not feasible to use raw datasets directly as summary statistics because the data typically lack meaningful ordering, since each observation is assumed to come from the same underlying process, making them statistically similar. However, \citet{bernton2019approximate} propose to use the Wasserstein distance as the distance metric in the ABC algorithm to directly compare empirical distributions of observed and simulated data, bypassing summary statistics selection. 

Researchers have also integrated ABC with existing sampling to improve its scalability and performance in high-dimensional parameter spaces. Several samplers have been developed, including ABC-MCMC \citep{marjoram2003markov, wegmann2009efficient, kypraios2017tutorial}, ABC-SMC \citep{sisson2007sequential, toni2009approximate, beaumont2009adaptive, drovandi2011likelihood}, and particle-based approaches such as ABC-pMCMC \citep{mckinley2020efficient}. These methods enable efficient sampling in high-dimensional parameter spaces, making ABC-class methods applied in epidemiology to infer parameters in transmission models and evaluate intervention strategies; see work of \citet{kypraios2017tutorial, gozzi2021estimating, gozzi2021importance, syga2021inferring,dankwa2022stochastic, gozzi2023estimating}. While these samplers can approximate the true posterior, they remain computationally intensive in models with complex latent structures or when real-time inference is required. As such, striking a balance between statistical accuracy and computational feasibility continues to be a central challenge in the development and application of ABC method.

\subsubsection{The Challenge of Summary Statistics Selection}

In ABC, summary statistics play an important role in enabling likelihood-free inference by simplifying high-dimensional data into low-dimensional representations. In the context of infectious disease modelling, these statistics often aim to capture critical epidemiological features such as the timing and magnitude of epidemic peaks, the total number of infections, or transmission dynamics across populations. However, selecting summary statistics that are both informative and computationally efficient remains one of the most significant challenges in the ABC framework. On the one hand, high-dimensional summary statistics can exacerbate the curse of dimensionality, leading to increased computational costs and reduced acceptance rates. On the other hand, overly simplistic summaries may discard essential information, resulting in poor posterior approximations \citep{ blum2010approximate}.

Recent machine learning approaches have begun to transform how summary statistics are constructed and selected in ABC. These data-driven methods aim to learn the functional relationship between data and parameters directly, thereby minimizing the reliance on expert-crafted statistics. \citet{jiang2017learning} used deep neural networks to fit the relationship between summary statistics and synthetic data. But deep neural networks need a lot of data to learn the parameters, so a three-layer neural network is used to run the model. \citet{raynal2019abc} introduce a method for automatically selecting informative summary statistics. This approach begins with a broad set of candidate statistics and uses random forests as a black-box regression tool to estimate posterior quantities based on these summaries. Moreover, \citet{aakesson2021convolutional} introduces a convolutional neural network architecture to directly map high-dimensional time-series data to parameter estimates, and the trained network's output (the predicted posterior means) serves as the summary statistics for the ABC inference process. For a detailed overview of additional methods for summary statistics selection, see Appendix B.

Overall, the integration of machine learning into ABC holds considerable promise for addressing longstanding challenges in summary statistic selection. Nonetheless, translating these methodological advances into robust and interpretable tools for real-world applications remains an open and critical area of ongoing research.

\subsection{Bayesian Synthetic Likelihood}

Bayesian Synthetic Likelihood (BSL) offers an alternative framework for performing approximate Bayesian inference when the likelihood is intractable. This section explores the BSL framework in detail.

\subsubsection{BSL Overview}

\begin{figure}
    \centering
    \includegraphics[width=1\linewidth]{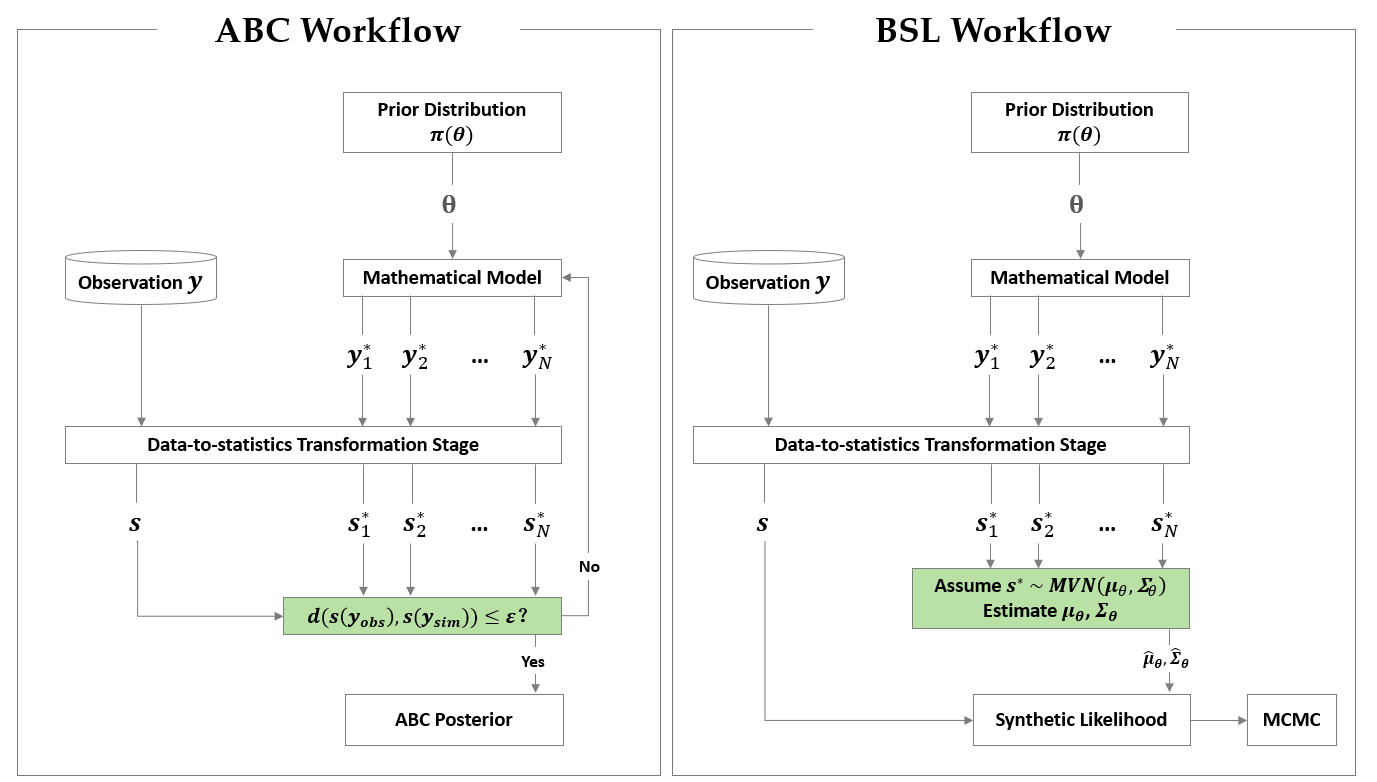}
    \caption{\textbf{Workflow comparison between Approximate Bayesian Computation (ABC) and Bayesian Synthetic Likelihood (BSL).} Both methods begin with prior sampling and simulation from a mathematical model to generate synthetic datasets ($y_1^*, y_2^*,...,y_N^*$). In ABC (left), observed and simulated datasets are transformed into summary statistics, and inference is based on whether the distance between them falls below a tolerance threshold, $\epsilon$. In contrast, BSL (right) assumes the simulated summary statistics follow a multivariate normal distribution, from which a synthetic likelihood is estimated and used within an MCMC framework. The key difference lies in how these methods handle summary statistics: ABC uses distance metric, while BSL constructs a synthetic likelihood.}
    \label{fig:abcbsl}
\end{figure}

The Synthetic Likelihood (SL) method, first introduced by \citet{wood2010statistical}, provides a simulation-based alternative to ABC method for addressing intractable likelihood problems. While both ABC and SL methods rely on data simulations, SL differs by assuming that the summary statistics, conditional on the model parameters, follow a multivariate normal distribution \citep{tong2012multivariate}. This assumption allows the likelihood to be expressed in terms of the unknown mean vector $\mu(\theta)$ and covariance matrix $\Sigma(\theta)$ of the summary statistics, which are estimated empirically through simulations. The workflow of the SL method (Figure \ref{fig:abcbsl}) involves the following steps:
\begin{enumerate}
    \item \textbf{Summary Statistics Reduction}: Reduce the observed data $y$ to a set of summary statistics $s$ to capture the dynamic structure of the model. Assume the summary statistics $s \sim N(\mu_\theta, \Sigma_\theta)$
    \item \textbf{Simulate Data}: Sample N parameter values $\theta_1,\theta_2,...,\theta_N$ from the prior distribution $\pi(\theta)$.
                \item \textbf{Generate Synthetic Data}: Simulate $N$ synthetic data sets, $y_1^*,y_2^*,...,y_N^*$ from the model given parameter $\theta$.
                \item \textbf{Compute Synthetic Summary Statistics}: Reduce the synthetic data sets to corresponding synthetic summary statistics vectors, $s_1^*,s_2^*,...,s_N^*$.
                \item \textbf{Estimate Unknown Parameters}:
$$\hat{\mu_\theta}=\sum \frac{s_i^*}{N}$$
$$\hat{\Sigma_\theta}=\frac{SS^T}{N-1}$$
where $$S=(s_1^*-\hat{\mu_\theta},s_2^*-\hat{\mu_\theta},...,s_N^*-\hat{\mu_\theta})$$
    \item \textbf{Construct Log- Synthetic Likelihood}:
$$l_s(\theta)=-\frac{1}{2}(s(y)-\hat{\mu_\theta})^T\hat{\Sigma_\theta}^{-1}(s(y)-\hat{\mu_\theta})-\frac{1}{2}log|\hat{\Sigma_\theta}|$$
\end{enumerate}
This synthetic likelihood can be directly optimized or incorporated into a Bayesian framework to form the Bayesian Synthetic Likelihood (BSL) posterior \citep{price2018bayesian}:
$$
\pi_N(\theta|s_{obs}) \propto N(s_{obs};\mu_N(\theta), \Sigma_N(\theta))\pi(\theta),
$$
where the subscript $N$ denotes the dependence on the number of simulations. Large numbers of simulations (i.e. large values of $N$) reduce likelihood variance but increase per-iteration costs. Conversely, smaller numbers of simulations lower costs but risk reduced acceptance rates due to higher variance in synthetic likelihood estimates. Therefore, it is important to find a balance between computational cost and likelihood variance.

\subsubsection{Methodological Advancement and Application}

Two major methodological challenges continue to shape the development of BSL: (1) reducing the computational burden associated with synthetic likelihood estimation; (2) selecting summary statistics that are both informative and compatible with the Gaussian assumption. Substantial methodological advancements have been made in both areas.

Synthetic likelihood estimation requires approximating the mean and covariance of summary statistics, which are unknown and must be estimated for each candidate parameter value. In the standard approach, these quantities are re-estimated at every iteration, leading to significant computational overhead, particularly as the dimensionality of the summary statistics increases. To mitigate this, several more efficient alternatives have been proposed. \citet{meeds2014gps} introduced a method that reduces estimation variance by employing a Gaussian process model for each parameter function. To simplify covariance estimation, their approach approximates the covariance matrix by modeling only its diagonal elements. In response, \citet{everitt2017bootstrapped} proposed a bootstrap-based covariance estimator, which empirically produces covariance estimates with lower variance compared to those obtained from raw sample estimates. Further improvements by \citet{an2019accelerating} use the Graphical Lasso to yield sparse, low-variance covariance estimates efficiently. Shrinkage estimators for the covariance matrix of summary statistics also contribute to reducing simulation demands \citep{ong2018likelihood, an2019accelerating}. Most recently, \citet{priddle2022efficient} proposed a method to decorrelate summary statistics using shrinkage with whitening transformation, resulting in improvements in computational efficiency. 

Nonetheless, a fundamental limitation of BSL persists: its reliance on the assumption that summary statistics are normally distributed. When this assumption is violated, the method's robustness can be compromised. In response, recent research has focused on relaxing the Gaussian constraint. \citet{fasiolo2018extended} introduced extended saddle point approximations as a more flexible likelihood approximation. \citet{thomas2022likelihood} proposed logistic regression-based synthetic likelihood methods to improve model adaptability. \citet{an2020robust} proposed a semiparametric density estimation framework that combines flexible marginal distributions with a Gaussian copula to account for non-normal dependence structures. Complementing these efforts, \citet{munoz2022combination} considered Poisson distributions for aggregated count data in epidemiological study. While these advances provide promising alternatives, their effectiveness depends critically on the structure of the chosen summary statistics.

\subsubsection{Application in Epidemiology}

BSL has been shown in many theoretical studies to be more tolerant of a higher-dimensional summary statistic than ABC \citep{price2018bayesian, frazier2021robust, frazier2023bayesian}. For example, \citet{price2018bayesian} showed that BSL significantly outperformed ABC in cell biology applications involving 145 summary statistics, where ABC struggled to reduce dimensionality without losing critical information contained in the image sequences.

Despite its theoretical advantages and methodological advancements, the application of BSL in epidemiology remains relatively limited. \citet{fasiolo2014statistical} applied BSL to state-space models in ecology and epidemiology. \citet{woroszylo2018modeling} provided one of the first applications of synthetic likelihood to real-world observational data, modeling household-level occurrences of diarrhea. More recently, \citet{munoz2022combination} incorporated BSL into a broader Bayesian framework to estimate parameters of a complex, large-scale epidemiological model.

As noted by \citet{drovandi2022comparison}, investing effort in selecting informative summary statistics can yield substantial improvements in model performance, often outperform full-data approaches. This observation highlights a valuable research direction of expanding the application of ABC and BSL methods to real-world epidemiological cases to further assess and validate their practical effectiveness.

\subsection{Integrated Nested Laplace Approximation}

Having examined approximate Bayesian inference methods, such as ABC and BSL, for models with intractable likelihoods, we now turn to approaches designed for models with tractable likelihoods. In particular, we focus on the Integrated Nested Laplace Approximation (INLA), a method specifically developed for efficient inference in Latent Gaussian Models (LGMs) \citet{rue2009approximate}.
 
\subsubsection{Latent Gaussian Models}

LGMs are a broad class of hierarchical models where the latent field is assumed to follow a Gaussian distribution, and the observations are conditionally independent given the latent field and hyperparameters. This class of models includes many commonly used statistical models, including generalized linear mixed models, spatial and spatio-temporal models based on Gaussian Markov random fields, and survival models, making INLA a practical tool for applied researchers \citep{rue2009approximate, rue2017bayesian}.

LGMs are structured hierarchically, consisting of three main components \citep{rue2009approximate}.

\textit{Stage 1: Observation model}

The observations $y_i$ are modeled as conditionally independent given the latent field $x$ and hyperparameters $\theta_1$. The likelihood can be expressed as:
$$
\textbf{y}|\textbf{x},\theta_1 \sim \prod\pi(y_i|x_i,\theta_1),
$$
where $x_i$ is the latent variable associated with the i-th observation.

\textit{Stage 2: Latent field}

The latent field $\textbf{x}$ is assumed to follow a Gaussian distribution with mean $\mu(\theta_2)$ and precision matrix $Q(\theta_2)$:
$$
\textbf{x}|\theta_2 \sim N(\mu (\theta_2), Q^{-1}(\theta_2))
$$
The precision matrix $Q(\theta_2)$ is often sparse, reflecting conditional independence properties in the latent field. In spatial models, for instance, this implies that disease risks in two regions are directly correlated only if they are neighbors, once the values in neighboring regions are known, the risks become conditionally independent. This dependency structure results in a sparse precision matrix, one that contains mostly zeros. Such sparsity is important for computational efficiency, as it allows for fast matrix operations \citep{rue2005gaussian}.

\textit{Stage 3: Hyperparameters}

The hyperparameters $\theta = (\theta_1, \theta_2)$ control the bahavior of latent field and/or the likelihood. A prior distribution $\pi(\theta)$ is assigned to these hyperparameters.

Key assumptions of LGMs include a small number of hyperparameters, a Gaussian latent field (often modeled as a Gaussian Markov Random Field, or GMRF) \citep{rue2005gaussian, held2010conditional}, and conditional independence of observations given the latent field and hyperparameters. These assumptions enable efficient computation and accurate approximations using INLA \citep{rue2009approximate}.

\subsubsection{INLA Methodology}

INLA provides a computationally efficient alternative to simulation-based methods like MCMC for approximating the posterior marginals of the latent field and hyperparameters in LGMs \citep{rue2017bayesian}. The main idea behind INLA is the use of nested Laplace approximations to approximate the high-dimensional integrals required to conduct Bayesian inference \citep{rue2009approximate}. The method can be summarized in three main steps:

Step 1: Approximation of the joint posterior of hyperparameters: The joint posterior $\pi(\theta|y)$ is approximated using a Laplace approximation \citep{barndorff1989asymptotic}:
$$
\tilde{\pi}(\theta|y) \propto \frac{\pi(x, \theta, y)}{\pi_G(x|\theta, y)}|_{x=x^*(\theta)},
$$
where $\pi_G(x|\theta,y)$ is a Gaussian approximation to the full conditional of $x$, and $x^*(\theta)$ is the mode of the full conditional for $x$ given $\theta$ \citep{rue2009approximate}.

Step 2: Approximation of the posterior marginals of the latent field: The posterior marginals $\pi(x_i|y)$ are approximated by integrating over the hyperparameters:
$$
\tilde{\pi}(x_i|y) = \int \tilde{\pi}(x_i|\theta, y)\tilde{\pi}(\theta|y)d\theta
$$
where $\tilde{\pi}(x_i|\theta,y)$ is approximated using either a Gaussian, Laplace, or simplified Laplace approximation \citep{rue2009approximate,martins2013bayesian}.

Step 3: Numerical integration over hyperparameters: The integration over the hyperparameters is performed using a grid or central composite design (CCD) strategy \citep{box1987empirical}, which allows for efficient exploration of the hyperparameter space \citep{rue2009approximate}.

INLA has been widely adopted due to its ability to provide fast and accurate approximations to posterior marginals, often outperforming MCMC in terms of computational efficiency, particularly for large datasets and complex models \citep{rue2017bayesian}. 

\subsubsection{Application in Epidemiology}

The R-INLA package offers a user-friendly interface for fitting LGMs using INLA, making it accessible to researchers in epidemiology \citep{martins2013bayesian}. 

Recent examples of applications include disease mapping and risk estimation. For instance, tools like SSTCDapp have been developed for estimating spatial and spatio-temporal disease risks using Bayesian hierarchical models \citep{adin2019online}. Researchers have also used INLA to compare discrete versus continuous spatial models for Bayesian disease mapping \citep{konstantinoudis2020discrete}, and to analyze the spatial distribution of HIV prevalence \citep{debusho2023bayesian}. 

During the COVID-19 pandemic, INLA played an important role in real-time spatio-temporal analysis, with tools like PandemonCAT tracking the pandemic's progression \citep{chaudhuri2022pandemoncat}.  It has been used to evaluate the impact of mobility restrictions \citep{saez2020effectiveness, jaya2023does}, estimate excess mortality \citep{knutson2023estimating}, and model SARS-CoV-2 reinfection dynamics \citep{penetra2023sars}. 

In infectious disease surveillance, INLA has been applied for correcting reporting delays \citep{bastos2019modelling}, improving decision-making for malaria intervention strategies \citep{canelas2021spatial, toh2021improving}, and modeling dengue transmission dynamics \citep{carabali2022joint, baldoquin2023potential}. 

For HIV and chronic disease epidemiology, INLA has improved the estimation of spatial heterogeneity in HIV risk groups \citep{wang2023msm, howes2023spatio}. 

Additionally, INLA has been applied to respiratory and environmental health, mapping respiratory infection risks \citep{cortes2023mapping} and analyzing spatial patterns in chronic obstructive pulmonary disease hospital admissions \citep{martinez2023spatial}. 

\subsubsection{Methodological Advancement}

Since its introduction by \citet{rue2009approximate}, INLA has revolutionized Bayesian inference for latent Gaussian models. Over the years, INLA has undergone significant methodological advancements, focusing on improving computational efficiency, extending the range of applicable models, and enhancing the accuracy and flexibility of the approximation process. 

One of the primary areas of advancement in INLA has been the improvement of computational efficiency. \citet{wang2022laplace} demonstrated the use of Laplace approximation within the Template Model Builder for the maximum likelihood estimation of intrinsic conditional autoregressive models. This approach significantly reduces computational time compared to MCMC methods and original INLA. Similarly, \citet{orozco2023big} propose a scalable divide-and-conquer approach for high-dimensional spatio-temporal disease mapping, partitioning spatial domains into smaller subdomains to reduce computational burden. \citet{van2023new} propose a modern reformulation of the INLA framework, which removes the linear predictors from the latent field, thereby reducing the computational burden associated with high-dimensional data. This reformulation, combined with a low-rank Variational Bayes correction, allows for faster inference without sacrificing accuracy.

Early implementations of INLA were primarily suited for latent Gaussian models with relatively simple structures. Recent methodological developments have broadened its applicability to more complex models. \citet{lee2022bayesian} introduce a Bayesian hierarchical modeling framework that uses penalized smoothing splines to create non-stationary spatial surfaces, allowing data-driven spatial structures - particularly useful for modeling infectious diseases with complex connectivity patterns. Additionally, \citet{jin2023epimix} 
propose EpiMix, a novel method for estimating the time-varying reproduction number for infectious diseases. EpiMix combines INLA with a Bayesian regression framework to incorporate the effects of exogenous factors and random effects. This integration allows for efficient estimation in real-time, even in low-incidence scenarios where traditional methods may struggle. The inlabru package \citep{bachl2019inlabru, lindgren2024inlabru} enables non-linear predictors and automates complex workflows, extended the capabilities of INLA to include point process models \citep{moller2007modern}, spatial count models, and distance sampling data \citep{miller2013spatial}, making advanced spatial modeling accessible to non-specialists. 

The development of the R-INLA package has played a central role in the adoption of INLA. \citet{van2021new} present several new developments in the INLA package that address the growing demand for scalable and efficient Bayesian inference in large-scale spatial, temporal, and spatio-temporal models.

Despite its many strengths, INLA is not universally applicable. Its reliance on LGMs restricts its use to cases where the latent field can be reasonably approximated as Gaussian. While extensions to non-Gaussian likelihoods have been successful, models with highly non-linear or non-Gaussian latent structures remain challenging to handle \citep{rue2009approximate}. For instance, non-Gaussian spatial effects often require additional approximations \citep{van2023new}, or alternative methods such as MCMC or hybrid approaches \citep{wang2022laplace}. Furthermore, the accuracy of INLA depends on the quality of the Laplace approximation, which may deteriorate in cases involving strong nonlinearities or high-dimensional random effects \citep{rue2009approximate}. 

Another limitation is INLA's performance in models with a large number of hyperparameters. While INLA excels for models with a moderate number of hyperparameters, its efficiency can degrade as the dimensionality of the hyperparameter space increases \citep{van2023new}. This is particularly relevant in complex spatio-temporal models or those with intricate interaction terms. Additionally, INLA struggles with certain hierarchical dependencies that MCMC handles naturally, and as model complexity grows, the numerical integration required in INLA becomes computationally intensive \citep{van2021new}.

Future research directions for INLA include the development of hybrid methodologies that integrate INLA with MCMC or variational inference techniques to better handle non-Gaussian latent structures \citep{van2021new, orozco2023big}. Advances in numerical integration techniques and adaptive algorithms could further improve INLA's scalability and accuracy for high-dimensional problems \citep{rue2009approximate}. Additionally, leveraging high-performance computing adaptations, such as GPU acceleration, could enhance INLA's salability for large-scale datasets \citep{van2021new}. Exploring connections between INLA and deep learning may also open new avenues for high-dimensional data analysis \citep{van2023new}.

\subsection{Variational Inference} 

In addition to the ABC, BSL, and INLA approaches to inference, Variational Inference (VI) is another influential approximate approach to constructing a likelihood in complex models.

\subsubsection{Overview of Variational Inference}

VI is an optimization-based approach for approximate Bayesian inference when the posterior distribution is analytically intractable or computationally expensive to sample \citep{jordan1999introduction, wainwright2008graphical}. Unlike MCMC, which samples a Markov chain and approximates the posterior with samples from the chain, VI transforms inference as an optimization problem. It tends to be faster and easier to scale to large data and is suitable for scenarios where quick exploration of many models is of interest \citep{blei2017variational}.

At its core, VI approximates the true posterior distribution $p(\theta|y)$ by selecting a tractable family of distributions $q(\theta;\phi)$ and optimizing its parameters $\phi$ to minimize their discrepancy from the posterior. This discrepancy is measured via the Kullback-Leibler (KL) divergence:
$$
\phi^*=arg \ min_{\phi \in \mathbb{\Phi}} \ \ KL(q(\theta;\phi) || p(\theta|y)),
$$
where the KL divergence quantifies the information lost (or entropy) when $q(\theta;\phi)$ replaces $p(\theta|y)$. Directly minimizing the KL divergence is infeasible because it requires evaluating the intractable posterior. Instead, VI maximizes the Evidence Lower Bound (ELBO), a surrogate objective derived from the KL divergence:
$$
ELBO(\phi) = E_{q(\theta)}[logp(y,\theta)]-E_{q(\theta)}[log q(\theta;\phi)].
$$
The ELBO can be interpreted as balancing two goals: (1) The term $E_{q(\theta)}[logp(y,\theta)]$ encourages the approximation to explain observed data well; and (2) The term $E_{q(\theta)}[log q(\theta;\phi)]$ penalizes deviations from the prior, preventing overfitting.

This optimization framework guarantees convergence and is inherently parrallelizable, possible for efficient inference even for large-scale models without compromising on the model complexity with the use of mean-field variational inference and stochastic variation inference \citep{hoffman2013stochastic}. 

\subsubsection{Applications in Epidemiology}

VI has become a powerful tool for addressing key challenges in modern epidemiological analysis, including real-time disease tracking, outbreak forecasting, and large-scale genomic surveillance.

VI has been used to rapidly infer time-varying epidemiological parameters from complex, noisy data streams. For instance, \citet{fan2016unifying} used VI with sigmoid belief networks to estimate latent infection states in a hierarchical graph-based Hidden Markov Model, capturing influenza spread over dynamic social networks without restrictive assumptions about low infection rate. During COVID-19, \citet{chen2021bayesian} applied Stein variational inference method to efficiently estimate high-dimensional, time-varying parameters in a heterogeneous COVID-19 epidemic model, providing uncertainty-quantified forecasts for long-term care facilities versus general populations. Further, \citet{wilder2021tracking} presented a Gaussian Process-based VI approach for estimating time-varying reproduction numbers from sparse and partially observed testing data. In agent-based modeling, \citet{smedemark2022probabilistic} applied black-box variational inference to estimate transmission parameters in network-driven SEIR simulations, allowing scalable inference from real world mobility and co-location data.

VI's ability to handle high-dimensional, sparse data has advanced disease forecasting. \citet{senanayake2016predicting} applied stochastic VI to scale Gaussian process regression for modeling and forecasting the spatio-temporal spread of seasonal influenza using large-scale epidemiological data. \citet{mcandrew2021adaptively} applied VI within a Bayesian ensemble framework, dynamically weighting models for real-time influenza forecasting under noisy and evolving surveillance data. Additionally, \citet{tahir2023prediction} introduced a Bayesian neural network using VI and flow normalization to predict the T-cell epitope response across major SARS-CoV-2 variants, generalizing across vaccinated and unvaccinated groups with imbalanced datasets.

In genomic epidemiology, VI has enabled phylodynamic analyses at scales that would be computationally infeasible for MCMC methods \citep{hassler2023data}. For example, \citet{ki2022variational} applied VI to estimate time-resolved effective reproduction number from hundreds of thousands of SARS-CoV-2 genomes in real time. Methods like those of \citet{moretti2021variational} and \citet{fourment2023automatic} have further shown VI's utility in optimizing phylogenetic tree topologies and handling gradient computations in phylogenetic models. However, despite its computational advantages, VI's application in Bayesian phylogenetics remains constrained by mean-field assumptions, intractable likelihoods, and poor scalability beyond modest tree sizes. Moreover, VI struggles to explore tree space effectively, often requiring topologists to guide the inference process \citep{fisher2022scalable}.

\subsubsection{Methodological Advancement}

VI has undergone a remarkable transformation in recent years, evolving from specialized, model-specific implementations to a versatile toolkit for scalable Bayesian computation \citep{blei2017variational}. This process has been driven by innovations in optimization techniques, flexible posterior approximations, and computational efficiency. While some of these advancements have yet to be widely adopted in infectious disease modeling, they have great potential to address key challenges in modern epidemiological research, such as high-dimensional parameter spaces, time varying latent states, and large-scale data.

Early VI approaches, such as mean-field approximations which assume that each parameter in the model can be updated independently and ignoring correlations, required labor-intensive derivations and struggled to scale to complex, high-dimensional models. \citet{kucukelbir2017automatic} introduced Automatic Differentiation Variational Inference (ADVI), which reframes VI as a generic optimization problem solvable via stochastic gradient ascent. By transforming constrained latent variables into an unconstrained real coordinate space and using automatic differentiation to compute gradients, ADVI allows scalable and black-box inference for a broad class of differentiable probabilistic models. It is deployed in the Stan probabilistic programming system \citep{kucukelbir2015automatic}, making it highly accessible for applied work. In Stan, the model gets defined using its domain-specific syntax, and users can switch from HMC to VI simply by setting a command-line flag. However, ADVI's reliance on Gaussian approximations can be restrictive for skewed and multimodal posteriors.
While VI in principle allows for more flexible, non-Gaussian variational families, such alternatives are rarely used in practice. Most implementations default to Gaussian approximations for analytical and computational convenience, making them restrictive in real-world applications, even though the underlying theory is capable of supporting further/diverse approximations.

Many infectious disease models involve latent processes that evolve dynamically, such as transmission rates, stochastic epidemic trajectories, or spatially correlated incidence patterns. Traditional VI struggles with such settings due to: (1) path dependencies means current states depend on their entire history (such as how today's case counts depend on past transmission), and (2) nonlinearities create complex relationships, such as threshold effects in herd immunity, both of which break VI's typical mean-field assumptions that treat variables as independent. To address this, \citet{ryder2018black} introduced a black-box VI framework for stochastic differential equations. Their approach uses a recurrent neural network to approximate the conditioned diffusion path, which avoids complex manual tuning and can effectively approximates complex conditional dynamics and supports fast, general purpose inference across a wide class of SDE models. For approximating complex hierarchical posteriors (e.g., local vs regional transmission), standard VI often collapses to a single mode or producing high-variance gradients. Nested Variational Inference \citep{zimmermann2021nested} addresses this by progressively refining the approximation through multiple layers: each level optimizes a KL divergence to learn intermediate target distributions. This method reduces gradient variance and improves sample quality in models with deep latent structures.

To improve computational efficiency in large-scale models, \citet{tan2018gaussian} proposed a Gaussian VI method that parameterizes the Cholesky factor of the precision matrix instead of the covariance matrix, allowing the incorporation of sparsity structures that reflect conditional independence structures in posterior distributions, and significantly reduce computational complexity in high-dimensional models. The method shows improved efficiency and accuracy on Generalized Linear Mixed Models and State-Space Models. Additionally, \citet{loaiza2022fast} introduces a hybrid VI approach that combines a flexible parametric approximation for global parameters with exact or approximate conditional sampling of latent variables. This approach retains MCMC style accurate posterior inference in models with many latent variables while keeping computation scalable.

Some newer VI methods leverage optimization dynamics for posterior approximation. Pathfinder \citep{zhang2022pathfinder} used quasi-Newton optimization paths and inverse Hessian-based local Gaussian approximations to efficiently locate high-probability regions of the posterior. By evaluating multiple paths in parallel and resampling via importance weighting, Pathfinder achieved scalable, accurate inference with significantly reduced computational cost compared to traditional VI. In addition, as deep learning gains traction in epidemiology, VI also supports uncertainty quantification in neural networks. \citet{chang2019ensemble} introduced ensemble-based VI methods that apply Gaussian mixtures selectively to critical network weights, allowing scalable Bayesian deep learning without prohibitive overhead. It is potentially useful for forecasting models or surrogate epidemic models using neural networks, where uncertainty estimation is needed but full Bayesian deep learning is impractical.

Despite its advancement, VI has limitations. A primary concern is the approximation bias introduced by the choice of restrictive variational family, particularly for heavy-tailed or multimodal distributions \citep{talts2018validating}. Unlike MCMC, VI lacks convergence diagnostics, complicating model validation \citet{gunapati2022variational}. Future research could develop more expressive but tractable variational families to better capture complex posterior distributions \citep{kucukelbir2017automatic}, and speed up computation for real-time surveillance while addressing robustness to model misspecification. Integrating VI with other inference methods (e.g., hybrid VI-MCMC) may combine the strengths of both frameworks \citep{salimans2015markov}.

\subsection{Comparison of Methods}

So far we have reviewed the methodological advancements of four families of approximate Bayesian inference methods and their applications within epidemiological research. Notably, INLA has been widely used, largely due to the accessibility and maturity of the R-INLA package. While ABC and VI have also been employed in epidemiological contexts. In many cases, applications are inspired by prior methodological studies or facilitated by the availability of software tools. For instance, ABC-rejection and ABC-SMC methods have been more frequently adopted following the work of \citet{minter2019approximate}, while ADVI \citep{kucukelbir2015automatic, kucukelbir2017automatic, chatzilena2019contemporary} has gained traction due to its implementation in the Stan platform. What accounts for this pattern of adoption? We argue that a comparison of the key features of these methods is warranted to better understand their respective advantages, limitations, and practical uptake (Table \ref{tab:compare_feature}).

\begin{table}
    \caption{\textbf{Comparison of Bayesian inference methods across key features.} This table summarizes five approaches we reviewed: Markov Chain Monte Carlo (MCMC), Approximate Bayesian Computation (ABC), Bayesian Synthetic Likelihood (BSL), Integrated Nested Laplace Approximation (INLA), and Variational Inference (VI), comparing their assumptions, requirements for parameter sampling and data simulation, methods for likelihood evaluation, and the nature of their posterior approximations.}
    \label{tab:compare_feature}
    \centering
    \includegraphics[width=1\linewidth]{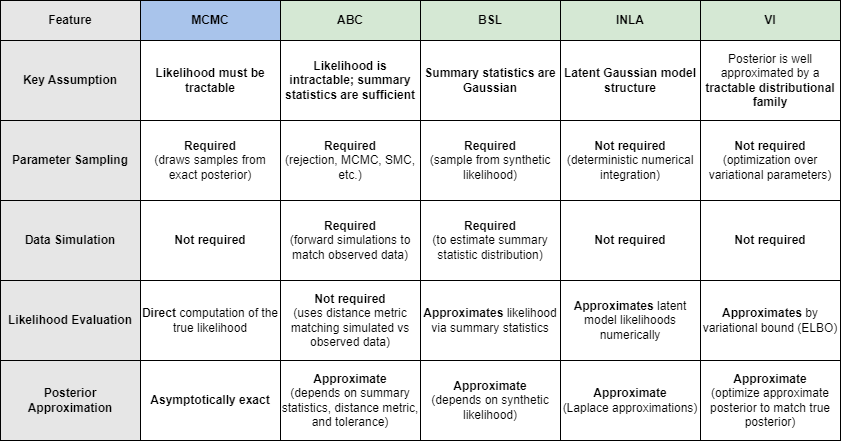}
\end{table}

MCMC methods offer asymptotically exact inference, making them a benchmark for accuracy in Bayesian analysis. Yet, they come with practical limitations. MCMC depends on the ability to compute the model's likelihood, and its computational cost can become prohibitive in complex, high-dimensional models. ABC provided a likelihood-free alternative by comparing summary statistics derived from observed and simulated data. While ABC is flexible and broadly applicable, it relies heavily on three key tuning parameters: summary statistics, distance metric, and tolerance. These dependencies, along with the need for repeated simulations, can lead to high computational demands. The BSL approach addresses some of these issues by modeling the distribution of summary statistics as Gaussian. This reduces computational costs compared to ABC. However, this assumption can limit its accuracy and applicability in systems that exhibit strong nonlinearity or multimodal posterior distributions. 

Unlike ABC and BSL, which rely heavily on simulations and carefully chosen summary statistics, INLA uses a combination of analytical and numerical integration to approximate posterior distributions deterministically. This makes it especially appealing for epidemiological models with hierarchical or spatial structures. That said, INLA's strengths come with a limitation: it is tailored for LGMs, and its use is best suited to cases where the latent process can reasonably be modeled as Gaussian. Variational Inference (VI) presents another scalable alternative by approximating the posterior distribution with a parametric family. While VI is computationally efficient and well-suited for high-dimensional problems, it introduces bias due to its reliance on specific variational families. Flexible variational families can better approximate complex posteriors but often at the cost of increased computational burden. Conversely, simpler families enhance efficiency but risk higher approximation error. Navigating this trade-off often requires accepting some degree of bias in exchange of speed and scalability. The inherent trade-offs across these methods, between accuracy, scalability, and robustness, motivate the growing interest in hybrid inference frameworks.

\begin{figure}
    \centering
    \includegraphics[width=1\linewidth]{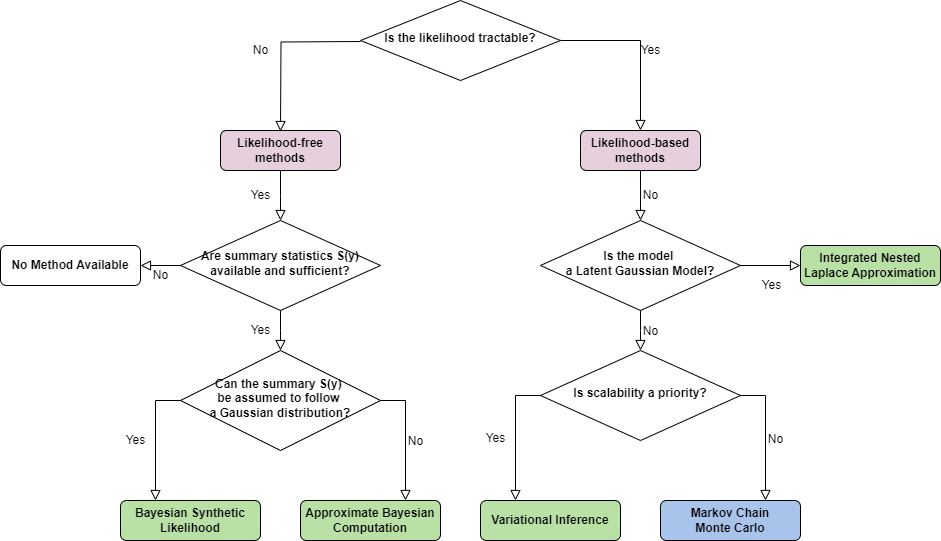}
    \caption{\textbf{Decision tree for selecting Bayesian inference methods in epidemiological modeling.}The flowchart guides the choice between and within likelihood-based and likelihood-free approaches based on key model characteristics.}
    \label{fig:decision_map}
\end{figure}

To enhance clarity and practical guidance for real-world applications, we have also developed a decision map (Figure \ref{fig:decision_map}) to aid in selecting the most suitable tools from the Bayesian inference toolbox. This map is structured around a series of key diagnostic questions designed to align methodological choices with the specific characteristics and demands of the research question at hand. The first step considers whether the likelihood function is tractable. If it is not, we turn to likelihood-free approaches. Within this branch, if informative and sufficient summary statistics are available, we further assess whether they follow a Gaussian distribution. If they do, BSL is appropriate; if not, ABC is the preferred method. If the likelihood function is tractable, we proceed with likelihood-based approaches. In this case, if the model structure aligns with a latent Gaussian model, INLA is well-suited. If the model is not a latent Gaussian model, we then consider whether scalability is priority. When scalability is critical, such as in high-dimensional or data-rich settings, VI is the preferred choice. Otherwise, MCMC remains the method of choice for its inferential accuracy.

\section{Conclusion}

In this paper, we have reviewed recent advances in both asymptotically exact and approximate Bayesian inference methods for infectious disease modeling. We have compared their key features to better understand their respective strengths, limitations, and practical uptake. To further support real-world application, we developed a decision map designed to guide the selection of appropriate tools from the Bayesian inference toolbox.

Exact Bayesian methods, such as MCMC, offer theoretically grounded inference but are often restricted by computational demands, the need for tractable likelihood functions, and the challenges posed by high-dimensional parameter spaces or partially observed epidemiological data. Approximate Bayesian inference methods have emerged as powerful alternatives that balance computational efficiency with reasonable inferential accuracy. Approaches such as ABC, BSL, INLA, and VI have significantly broadened the scope for rapid analysis of complex models, particularly when exact methods are computationally prohibitive.

Nonetheless, approximate methods are not without limitations. Approximation biases and the lack of robust diagnostics for posterior accuracy remain persistent challenges. These challenges highlight two research frontiers: (1) developing hybrid Bayesian inference methods that strategically integrate the strengths of both exact and approximate techniques, to achieve scalable yet theoretically grounded inference; and (2) applying these advances to answer important epidemiological questions, meeting the growing need for accurate and efficient inference during public health crises.
Progress in these areas will also be accelerated by developments in machine learning, probabilistic programming, and automated tuning strategies \citep{vstrumbelj2024past}, opening rich interdisciplinary opportunities for collaboration.

\section{Acknowledgments}

The authors thank the University of St Andrews and the China Scholarship Council for their financial support, which made this research possible. We are also grateful to the Centre for Research on Ecological and Environmental Modelling (CREEM) for providing the resources and environment necessary for the successful completion of this work. We also appreciate the constructive feedback and insights provided by my colleagues and the research community at CREEM.

\bibliographystyle{apalike}
\bibliography{main}

\newpage
\section*{Appendices}
\addcontentsline{toc}{section}{Appendices}

\subsection*{Appendix A: Hamiltonian Monte Carlo}

The HMC algorithm proceeds through the following key steps:

\begin{enumerate}
    \item \textbf{Initialization}: 
    Define the joint density of the target parameters $\theta$ and auxiliary momentum variables $r$, typically chosen as: 
    $$
    \pi(\theta, r) = \pi(\theta) \cdot N(r|0, I),
    $$
    where $r$ is sampled from a Gaussian distribution. This joint density combines the posterior distribution of $\theta$ (potential energy) with the momentum distribution of $r$ (kinetic energy) to define the total energy of the system. 
    \item \textbf{Leapfrog Integration}: 
    Simulate the system's Hamiltonian dynamics using the following differential equations:
    \[
    \frac{d\theta}{dt} = \frac{\partial H}{\partial r}, \quad \frac{dr}{dt} = -\frac{\partial H}{\partial \theta}
    \]
    where the Hamiltonian, $H(\theta, r) = -\log \pi(\theta, r)$, represents the total energy.  The leapfrog integrator alternates updates to position $(\theta)$ and momentum $(r)$, ensuring numerical stability and adherence to the system's dynamics \citep{beskos2013optimal}. This approach uses the gradients of the posterior to guide the trajectory, allowing the algorithm to propose informed candidate states efficienctly, even in high-dimensional parameter spaces.
    \item  \textbf{Metropolis Acceptance Step}: 
    Correct for numerical errors introduced by the leapfrog integrator by performing an acceptance-rejection step. The proposed state $(\theta',r')$ is accepted with probability:
    $$
    \alpha = \min \left(1, \frac{\pi(\theta', r')}{\pi(\theta, r)} \right),
    $$ ensuring the Markov chain asymptotically converges to the true posterior distribution.
\end{enumerate}

\begin{figure}
    \centering
    \includegraphics[width=1\linewidth]{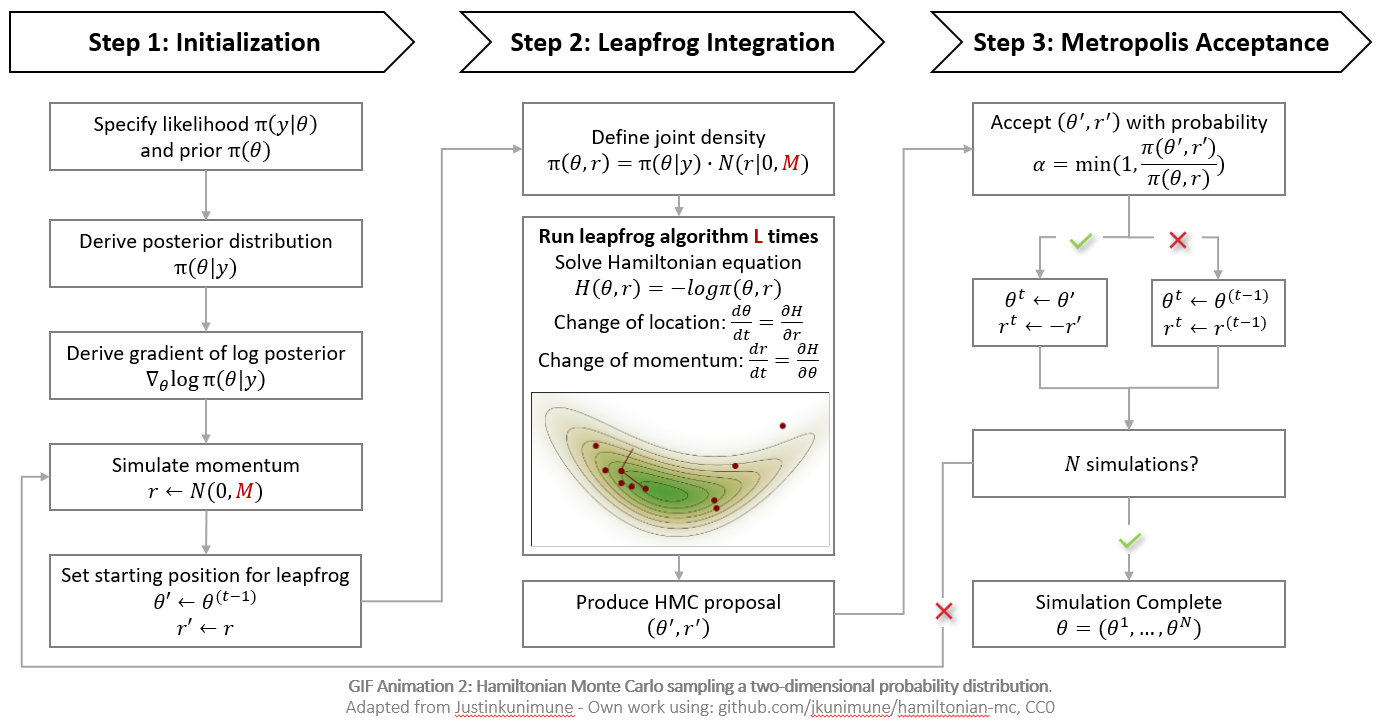}
    \caption{\textbf{Workflow of the HMC algorithm.} The HMC procedure consists of three main steps: (1) Initialization, where the posterior distribution and its gradient are derived, and a random momentum variable is simulated; (2) Leapfrog Integration, where the Hamiltonian dynamics are numerically solved using the leapfrog algorithm to propose a new state; and (3) Metropolis Acceptance, where the proposal state is accepted or rejected based on the Metropolis criterion. This process iterates until the desired number of posterior samples is generated.}
    \label{fig:hmc_workflow}
\end{figure}

A detailed HMC workflow can see Figure \ref{fig:hmc_workflow}.

\subsection*{Appendix B: Summary Statistics Selection Method for ABC}

As outlined by \citet{prangle2018summary}, summary statistic selection methods fall into three main categories: projection, auxiliary likelihood, and subset selection. Projection methods reduce dimensionality by transforming a set of high-dimensional candidate statistics into a lower-dimensional space tailored for inference. Techniques such as partial least squares regression \citep{wegmann2009efficient}, boosting \citep{aeschbacher2012novel}, and linear regression-based adjustments \citep{fearnhead2012constructing} fall into this category. While computationally efficient, projection methods often trade off interpretability, making it harder to link transformed features back to the original data/mechanisms. Auxiliary likelihood methods approximate the likelihood using auxiliary models, such as maximum likelihood estimators (MLEs) \citep{drovandi2011approximate,wilson2009rapid, gleim2013approximate}, or scores \citep{gleim2013approximate}.  These methods bypass the need to predefine candidate statistics and are particularly advantageous in cases where an auxiliary likelihood is well-aligned with the true data-generating process. However, their effectiveness heavily depends on the quality of the auxiliary model \citep{drovandi2015bayesian}, which may be difficult to specify in complex epidemic models. Subset selection methods aim to identify a low-dimensional, informative subset of statistics from a larger candidate pool by optimizing criteria such as entropy reduction \citep{Nunes2010} or mutual information \citep{joyce2008approximately}. These methods are valued for their interpretability, as they directly highlight which statistics are most informative. However, these methods assume the existence of an optimal low-dimensional subset, which may limit their applicability in highly complex models.

\end{document}